\documentclass{ifacconf}

\usepackage{graphicx}      
\usepackage{natbib}        
\usepackage{amsmath}
\usepackage{amssymb}
\usepackage{mathdots}
\usepackage{natbib}
\usepackage[ruled,vlined,algo2e]{algorithm2e}
\usepackage[all,cmtip]{xy}
\usepackage{tikz}
\usepackage{xcolor} 
\usetikzlibrary{automata,positioning,shapes,arrows}

\theoremstyle{definition}
\newtheorem{definition}{Definition}[section]
\theoremstyle{definition}
\newtheorem{notation}{Notation}[section]
\begin{document}
\begin{frontmatter}

\title{A Subspace Method for Time Series Anomaly Detection in Cyber-Physical Systems} 


\author[First]{Fredy Vides} 
\author[Second]{Esteban Segura} 
\author[Third]{Carlos Vargas-Ag\"{u}ero}

\address[First]{Scientific Computing Innovation Center, Universidad Nacional Aut\'onoma de Honduras, 
   Tegucigalpa, Honduras (e-mail: fredy.vides@unah.edu.hn).}
\address[Second]{Escuela de Matem\'atica - CIMPA, Universidad de Costa Rica, 
   San Jos\'e, Costa Rica (e-mail: esteban.seguraugalde@ucr.ac.cr)}
\address[Third]{Escuela de F\'isica, 
Universidad de Costa Rica, San Jos\'e, Costa Rica, (e-mail: carlos.vargasaguero@ucr.ac.cr)}

\begin{abstract}                
Time series anomaly detection is an important process for system monitoring and model switching, among other applications in cyber-physical systems. In this document, we present a fast subspace method for time series anomaly detection, with a relatively low computational cost, that has been designed for anomaly detection in real sensor signals corresponding to dynamical systems. We also present some general results corresponding to the theoretical foundations of our method, together with a prototypical algorithm to for time series anomaly detection. Some numerical examples corresponding to applications of the prototypical algorithm are presented, and some computational tools based on the theory and algorithms presented in this paper, are provided. 
\end{abstract}

\begin{keyword}
Anomaly detection, Hankel matrix, time series analysis, sensors, signals. 
\end{keyword}

\end{frontmatter}

\section{Introduction}
In this document, a subspace method for time series anomaly detection with applications to the computational analysis of sensor signals in cyber-physical systems, is presented.

As smart devices become more and more popular, in order to guarantee a smooth interaction between humans and the aforementioned devices, it is important to have flexible enough time series anomaly detection algorithms that can contribute to a suitable modeling and communication between humans and smart devices.

Some smart devices of particular interest for us, are wearable devices like {\em smart bands} that can be used for fitness and health tracking or monitoring, and are able to work independently or connected to the cloud via internet of things, as considered in \cite{Wearables1_ZHAO2021691} and \cite{Wearables2_Passos2021}. As these devices are equipped with several sensor sets that can be used to collect real-time signals corresponding to a given physical activity or mood state, and since they can be used alone or paired with smartphones or other devices, it is convenient to have anomaly detection methods with relatively low computational requirements, and even if one connects smart-bands to more powerful computers, having a fast low cost anomaly detection method may be convenient for research and daily life applications of wearable devices. 

Time series anomaly detection is also useful for data driven cyber-physical system identification in the sense of \cite{Yuan2019}, where an automatic modeling algorithm needs to be able to switch between models, in response to anomalies detected on signals corresponding to sensor measurements.

The main contributions of this work are: the presentation of a subspace numerical technique for time series anomaly detection, and a computational implementation written in Python. The subspace method for time series anomaly detection presented in this document can be used independently, or combined with other anomaly detection methods, using artificially intelligent schemes like the ones presented in \cite{FVides_SPAAR} to switch to more computationally complex methods, whenever there are enough resources to do so.

The rest of the document is organized as follows. In section \S\ref{sec:preliminaries} some preliminaries and general notation are introduced. The main theoretical results obtained as part of this study are presented in \S\ref{sec:anomaliesDetector}. A prototypical algorithm is presented in section \S\ref{sec:algorithms} and some numerical experiments are documented in \S\ref{sec:experiments}. Some hints and future directions are outlined in section \S\ref{sec:Future}.

\section{Preliminaries}\label{sec:preliminaries}
In this document, we will write $\mathbb{S}^{n-1}$ to denote the complex $(n-1)$-sphere defined as the set 
$$\mathbb{S}^{n-1} = \{ \mathbf{x} \in \mathbb{C}^n: \| \mathbf{x} \| = 1\}.$$ We will use the expression $\mathbf{I}_n$ to denote the identity matrix in $\mathbb{C}^{n\times n}$. 

Given a ${\displaystyle m\times n}$ complex matrix $\mathbf{M}$, the singular value decomposition (SVD) of $\mathbf{M}$ is a factorization of the form ${\displaystyle \mathbf {M} = \mathbf {U\Sigma V^{*}} }$, where $\mathbf{U}$ is an ${\displaystyle m\times m}$ complex unitary matrix, ${\displaystyle \mathbf {\Sigma } }$ is an ${\displaystyle m\times n}$ rectangular diagonal matrix with non-negative 
real numbers on the diagonal and $\mathbf{V}$ is an ${\displaystyle n\times n}$ 
complex unitary matrix. Moreover, the diagonal entries ${\displaystyle \sigma _{j}(\mathbf{M})=\Sigma _{jj}}$ of $ {\displaystyle \mathbf {\Sigma } }$, known as the singular values of $\mathbf{M}$, 
are uniquely determined by $\mathbf{M}$ and the columns of $\mathbf{U}$ and the 
columns of $\mathbf{V}$ are called left-singular vectors and right-singular vectors of $\mathbf{M}$, respectively (see, for instance, \cite{golub2013matrix}). 

As presented in \cite{demmel1997applied}, we have that the power method generates a sequence of approximate 
eigenvectors converging to an eigenvector associated to the largest (in absolute value) eigenvalue. In this work, we use a variation known as the inverse power method with 
Rayleigh quotient (IPM) to approximate an eigenvector corresponding to the smallest (in absolute value) eigenvalue. 
The idea of this method goes as follows: choose an initial approximation to the eigenvector $ \mathbf{u}_{0} \neq 0$ such that $\left\|\mathbf{u}_{0}\right\|_{2}=1$. Then, 
compute the initial approximation to the associated eigenvalue by: $\sigma_{0}= \mathbf{u}_{0}^{T} \mathbf{A} \mathbf{u}_{0}$. Afterwards, for each iterative step, $i$, we compute the new approximation to the 
eigenvector by $\mathbf{v}_{i+1}=\left(\mathbf{A}-\sigma_{i} \mathbf{I}_n\right)^{-1} \mathbf{u}_{i}$, we normalize this $\mathbf{u}_{i+1}=\mathbf{v}_{i+1} /\left\| \mathbf{v}_{i+1}\right\|_{2}$ and then we approximate the associated eigenvalue: $\sigma_{i+1}=\mathbf{u}_{i+1}^{T} \mathbf{A} \mathbf{u}_{i+1}$. In Algorithm \ref{alg:main_anomaly_alg_3} in \S\ref{sec:algorithms}, we use the inverse power method with Rayleigh quotient because it is known that  is cubically convergent for symmetric matrices. 

Now, given $\delta>0$, let us consider the function defined by the expression 
\begin{align*}
H_\delta(x)=\left\{
\begin{array}{ll}
1, & x>\delta\\
0,& x\leq \delta
\end{array}
\right..
\end{align*}

Consider a matrix $\mathbf{A}\in \mathbb{C}^{m\times n}$ with singular values denoted by $\sigma_j(\mathbf{A})$, for $j=1,\ldots,\min\{m,n\}$. Then, as presented in \citep{FVides_SPAAR}, we will write $\mathrm{rk}_\delta(\mathbf{A})$ to denote the nonnegative integer determined by the expression
\begin{equation*}
\mathrm{rk}_\delta(\mathbf{A})=\sum_{j=1}^{\min\{m,n\}}H_\delta(\sigma_j(\mathbf{A}))
\end{equation*}
On the other hand, remember that the kernel of $\mathbf{A}$ is the set of solutions to the equation $\mathbf{A}\mathbf{x} = \mathbf{0}$, where $\mathbf{0}$ is understood as the zero vector. Thus, as decribed in \cite{axler1997linear}, the kernel of $\mathbf{A}$ is defined as
$$
\operatorname{ker}(\mathbf{A}) = \left\{\mathbf{x} \in \mathbb{C}^{n} : \mathbf{A} \mathbf{x}=\mathbf{0}\right\}
$$

Given an positive integer $L$ and any time series $\Sigma_{\infty}=\{x_t\}_{t\geq 1}$, we will write $\mathbf{x}_L(t)$ to denote the vector in $\mathbb{R}^L$ determined by the expression:
\begin{equation}
    \mathbf{x}_L(t) = \begin{bmatrix}
    x_{t-L+1} \\ x_{t-L+2} \\ \vdots \\ x_{t}
    \end{bmatrix}
\label{eq:xL}
\end{equation}
for any $t\geq L$. Moreover, given a lag value $L>0$ and an ordered sample $\Sigma_{N}=\{x_{t}\}_{t=1}^N\subset \Sigma_\infty$ from a time series $\Sigma_\infty=\{\hat{x}_{t}\}_{t\geq 1}$, we will write $\mathcal{H}_L(\Sigma_N)$ to denote the Hankel-type trajectory matrix corresponding to $\Sigma_{N}$, that is determined by the following expression.
\begin{align*}
\mathcal{H}_L(\Sigma_N)&=\begin{bmatrix}
x_1 & x_2 & x_3 & \cdots & x_{N-L+1}\\
x_2 & x_3 & x_4 & \cdots & x_{N-L+2}\\
\vdots & \vdots & \vdots & \iddots & \vdots\\
x_{L} & x_{L+1} & x_{L+2} & \cdots & x_{N} 
\end{bmatrix}\nonumber\\
&=\begin{bmatrix}
| & | & & |\\
\mathbf{x}_{L}(L) & \mathbf{x}_{L}(L+1) & \cdots & \mathbf{x}_{L}(N)\\
| & | & & |
\end{bmatrix}
\label{eq:Hankel_v_rep}
\end{align*}

In this document, every time we write that $\Sigma\subset \Sigma_\infty$ is a time series sample from $\Sigma_\infty=\{x_t\}_{t\geq 1}$, we will be implicitly considering the particular case where $\Sigma=\{x_t\}_{t=1}^T$, for some positive integer $T$.

The column space of a matrix $\mathbf{A}\in \mathbb{C}^{m\times n}$ will be denoted by $\mathrm{colsp}(\mathbf{A})$ in this study. A matrix $\mathbf{P} \in \mathbb{C}^{n\times n}$ will be called an orthogonal projector whenever $\mathbf{P}^2=\mathbf{P}=\mathbf{P}^\ast$, where $\mathbf{P}^\ast$ denotes the conjugate transpose $\overline{\mathbf{P}}^\top$ of $\mathbf{P}$.

Given two (column) vectors $\mathbf{x}=[x_1 \; x_2 \; \cdots \; x_n]^\top$ and $\mathbf{y}=[y_1 \; y_2 \; \cdots \; y_n]^\top$ in $\mathbb{C}^n$, we will write $\mathbf{x}\odot \mathbf{y}$ to denote de Hadamard product determined by the operation
\begin{align*}
 \mathbf{x}\odot \mathbf{y}=\begin{bmatrix}
 x_1 y_1\\
 x_2 y_2\\
 \vdots\\
 x_ny_n
 \end{bmatrix}.   
\end{align*}

\section{Projective time series anomaly detection}
\label{sec:anomaliesDetector}

We will start this section by introducing what we call a $L$-normality region for a certain type of time series sample $\Sigma\subset \Sigma_\infty$ from some given time series $\Sigma_\infty=\{x_t\}_{t\geq 1}$, and some given lag value $L>0$. 

\begin{notation}\label{nota:normality-region}
Given a positive integer $L$ and a time series sample $\Sigma=\{x_t\}_{t=1}^T \subset \Sigma_\infty$ corresponding to the first $T\geq L+1$ elements of $\Sigma_\infty=\{x_t\}_{t\geq 1}$, we will denote by $\mathcal{N}_L(\Sigma)$ the set of all $\mathbf{x}_L(t)\in \mathrm{colsp}(\mathcal{H}_L(\Sigma))$. The set $\mathcal{N}_L(\Sigma)$ will be called the $L$-normality region for $\Sigma$.
\end{notation}

The notion of $L$-normality region for a time series sample previously considered, allows one to define the following concept.

\begin{definition} \label{def_PAD}
Given $\varepsilon\geq 0$, a lag value $L>0$, and a time series sample $\Sigma\subset \Sigma_\infty$ corresponding to the first $T\geq L+1$ elements of the time series $\Sigma_{\infty}=\{x_t\}_{t\geq 1}$. A $(L,\varepsilon)$ Projective Anomaly Detector ($(L,\varepsilon)$-PAD) based on $\Sigma$ for $\Sigma_\infty$ is determined by a projector $P$ such that:
$$
\| \mathbf{P} \mathbf{x}_L(t) - \mathbf{x}_L(t) \| \leq \varepsilon,
$$
when the $L$-component $x_t$ of $\mathbf{x}_{L}(t)$ does not have an anomalous value, with respect to the training data determined by the columns of $\mathcal{H}_L(\Sigma)$, that is, when $\mathbf{x}_L(t)\in \mathcal{N}_L(\Sigma)$. 
\end{definition}

Building on the notion of the kernel of a matrix, we can define the approximate kernel of a matrix $\mathbf{A}$ as follows. 

\begin{definition}  \label{approxKer}
Let $\varepsilon>0$. The approximate kernel of $\mathbf{A}$, denoted by $\operatorname{ker}_\varepsilon (\mathbf{A})$, can be defined as: 
$$
\operatorname{ker}_\varepsilon (\mathbf{A}) = \left\{\mathbf{x} \in \mathbb{S}^{n-1} : \| \mathbf{A} \mathbf{x} \| \leq \varepsilon \right\} \; \cup \; \{ \mathbf{0} \}.
$$
\end{definition}

From the definition of the approximate kernel, it can be seen that $\operatorname{ker}(\mathbf{A}) \subseteq \operatorname{ker}_\varepsilon(\mathbf{A})$. 

Based on the notions previously considered, we can obtain the following theorem. 

\begin{thm} \label{theo1}
Given a lag value $L>0$ and a time series sample $\Sigma\subset \Sigma_\infty$ with corresponding Hankel type trajectory matrix $\mathcal{H}_L(\Sigma)$. For any $\sigma_1(\mathcal{H}_L(\Sigma))\geq  \varepsilon \geq \sigma_L(\mathcal{H}_L(\Sigma))$, there is a projector $\mathbf{P}$ that satifies the condition
\begin{equation}
\| \mathbf{P} \mathbf{x}_{L}(t) - \mathbf{x}_{L}(t) \| \leq K_L(t)\varepsilon, \label{eq:weak-PAD}
\end{equation} 
for each $\mathbf{x}_{L}(t) \in \mathcal{N}_L(\Sigma)$ and some number $K_L(t)$ that does not depend on $\varepsilon$, for $t\geq L$.
\end{thm}

\begin{pf}
Let us denote by $M$ the number of columns of $\mathcal{H}_L(\Sigma)$. In order to prove the existence of the projector $\mathbf{P}$ that satisfies \eqref{eq:weak-PAD}, we can start by considering a vector $\mathbf{x}_{L}(t)$ determined by the elements of the time series $\Sigma_\infty=\{x_t\}_{t\geq 1}$ containing the time series sample $\Sigma$, according to Equation \eqref{eq:xL}, for $t\geq L$, such that $\mathbf{x}_L(t)\in \mathrm{colsp}(\mathcal{H}_L(\Sigma))$. Consequently, there is a solution $\mathbf{c}(t)=[c_1(t) \; \cdots \; c_M(t)]^\top$ to the linear system
\begin{equation}
\mathcal{H}_L(\Sigma) \mathbf{c}(t) = \mathbf{x}_{L}(t). 
\label{eq:LS-proof}
\end{equation}
From the SVD decomposition 
\begin{equation}
\begin{bmatrix}
| & | &  & |\\
\mathbf{u}_1 & \mathbf{u}_2 & \cdots & \mathbf{u}_L\\
| & | &  & |
\end{bmatrix}
\mathbf{\Sigma V}=\mathcal{H}_L(\Sigma),
\label{eq:proof-svd}
\end{equation}
let us consider the $r=\mathrm{rk}_\varepsilon(\mathcal{H}_L(\Sigma))$ left singular vectors $\mathbf{u}_1,\ldots,\mathbf{u}_r$ of $\mathcal{H}_L(\Sigma_N)$ determined by \eqref{eq:proof-svd}. From \eqref{eq:proof-svd} we will have that $\mathrm{span}(\{\mathbf{u}_1,\ldots,\mathbf{u}_L\})=\mathrm{colsp}(\mathcal{H}_L(\Sigma))$, and since \begin{equation}
\sigma_r(\mathcal{H}_L(\Sigma_N))\geq \varepsilon>\sigma_{r+1}(\mathcal{H}_L(\Sigma_N)),
\label{eq:sigma-sep}
\end{equation}
as a consequence of \cite[Theorem 3.4]{DBLP:journals/corr/abs-2105-07522} we will have that if we set  $s_r=\sqrt{r(\min\{L,M\}-r)}$, 
\[
\mathbf{U}_r:=\begin{bmatrix}
| & | &  & |\\
\mathbf{u}_1 & \mathbf{u}_2 & \cdots & \mathbf{u}_r\\
| & | &  & |
\end{bmatrix}
\]
and $\mathbf{P}:=\mathbf{U}_r\mathbf{U}_r^\ast$ then, 
\begin{equation}
\left\|\mathcal{H}_L(\Sigma)-\mathbf{P}\mathcal{H}_L(\Sigma)\right\|_F\leq \frac{s_r}{\sqrt{r}}\varepsilon.
\label{eq:first-PAD-estimate}
\end{equation}
Consequently, by \eqref{eq:LS-proof} and \eqref{eq:first-PAD-estimate} we will have that
\begin{align*}
\left\|\mathbf{P}\mathbf{x}_L(t)-\mathbf{x}_L(t)\right\|&\leq
\left\|\mathcal{H}_L(\Sigma)-\mathbf{P}\mathcal{H}_L(\Sigma)\right\|_F\|\mathbf{c}(t)\|\nonumber\\
&\leq \frac{s_r}{\sqrt{r}}\|\mathbf{c}(t)\|\varepsilon.
\end{align*}
Let us set 
\begin{equation}
    K_L(t):=\frac{s_r}{\sqrt{r}}\|\mathbf{c}(t)\|.
    \label{eq:Kt-def}
\end{equation}
This completes the proof.
\end{pf}

\begin{notation}
Given a positive integer $L$ and a time series sample $\Sigma\subset \Sigma_\infty$. For each $t\geq L$, the numbers $K_L(t)$ determined by \eqref{eq:Kt-def} will be called the $L$-relative bounds for the time series $\Sigma_\infty$ with respect to the sample $\Sigma$.
\end{notation}

Using the $L$-relative bounds for time series previously considered, we can obtain the following corollary.

\begin{cor}\label{cor:PAD-cor}
Given a lag value $L>0$, a positive integer $N>L$ and a time series sample $\Sigma\subset \Sigma_\infty$ with $L$-relative bounds $K_L(t)$, for $t\geq L$. If we set
\begin{equation}
    \hat{K}:=\sup_{L\leq t\leq N} K_L(t),
    \label{eq:KL_region}
\end{equation}
then there are $\varepsilon\geq 0$ and a $(L,\hat{K}\varepsilon)$-PAD $\mathbf{P}$ based on $\Sigma$ for $\Sigma_\infty$, for $L\leq t\leq N$.
\end{cor}
\begin{pf}
Let us set $\mathbf{H}:=\mathcal{H}_L(\Sigma)$. If we choose $\sigma_1(\mathbf{H})\leq \varepsilon\leq\sigma_L(\mathbf{H})$, then by Theorem \ref{theo1} we will have that there is a projector $\mathbf{P}$ such that 
\begin{equation}
\| \mathbf{P} \mathbf{x}_{L}(t) - \mathbf{x}_{L}(t) \| \leq K_L(t)\varepsilon \leq \hat{K}\varepsilon,\label{eq:almost-kernel}
\end{equation}
for each $\mathbf{x}_L(t)\in \mathcal{N}(\Sigma)$, for $L\leq t\leq N$. This completes the proof.
\end{pf}

The relation \eqref{eq:almost-kernel} obtained as part of the proof of Corollary \ref{cor:PAD-cor} allows one to observe the following.

\begin{rem}\label{rem:Almost-ker-rem}
Given a lag value $L>0$, a positive integer $N>L$ and a time series sample $\Sigma\subset \Sigma_\infty$. If we set
\begin{equation}
    \hat{K}:=\sup_{L\leq t\leq N} K_L(t),
    \label{eq:KL_region_2}
\end{equation}
then there are $\varepsilon\geq 0$ and  a projector $\mathbf{P}$ such that 
\begin{equation}
\|(\mathbf{I}_L - \mathbf{P}) \mathbf{x}_{L}(t)\| \leq  \hat{K}\varepsilon,\label{eq:almost-kernel-2}
\end{equation}
for each $\mathbf{x}_L(t)\in \mathcal{N}(\Sigma)$, for $L\leq t\leq N$. Consequently, every unit vector $\mathbf{q}\in \mathrm{colsp}(\mathbf{I}_L-\mathbf{P})=\mathrm{ker}(\mathbf{P})$ will belong to $\mathrm{ker}_{\nu(\varepsilon)}(\mathcal{H}_L(\Sigma)^\ast)$, for $\nu(\varepsilon)=\sqrt{N}\hat{K}\varepsilon$, where $N$ denotes the number of columns of $\mathcal{H}_L(\Sigma)$.
\end{rem}

\begin{figure}[h!]
	\centering
	\includegraphics[width=0.45\textwidth]{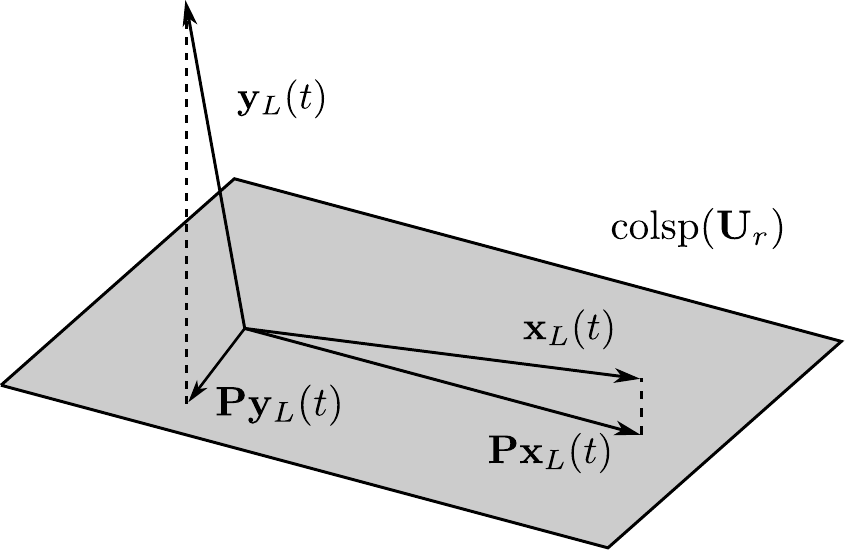}
	\caption{Representation of a {\em normal} vector signal $\mathbf{x}_{L}(t)$ and an {\em anomalous} vector signal $\mathbf{y}_{L}(t)$ projected on $\mathrm{colsp}(\mathbf{U}_r)$}
	\label{fig:subespacio}
\end{figure}

A heuristic approach the the previously considered concepts allows one to observe that, if there is an anomaly in the vector signal $\mathbf{x}_{L}(t)$, we should not be able to write it approximately as a linear combination
of the column vectors of $\mathbf{P}$, i.e $\|\mathbf{P}  \mathbf{x}_{L}(t) - \mathbf{x}_{L}(t) \|$ is not small. So the vector $\mathbf{x}_{L}(t)$ with anomalous components must have a significative component on the orthogonal complement of $\mathrm{colsp}(\mathbf{U}_r)$. We 
can measure the projection on the left singular vector $\mathbf{u}$ corresponding to the smallest singular value, and use it to determine if there are any anomalies. This is shown graphically in the Fig. \ref{fig:subespacio}, where $\mathbf{x}_{L}(t)$ is a {\em normal} vector signal and $\mathbf{y}_{L}(t)$ has some {\em anomalous components}. 
If we have random noise on the signal, the random noise is going to be distributed uniformly on all the left singular vectors \citep{meyer2000applied}. Considering the Fig. \ref{fig:subespacio}, the non anomalous signal $\mathbf{x}_{L}(t)$ with noise is not going to be really close to the subspace $\mathrm{colsp}(\mathbf{U}_r)$. If there is a truly anomalous behavior that is not random noise, the anomalous signal $\mathbf{y}_{L}(t)$ is going to have a strong projection in the orthogonal component of $\mathrm{colsp}(\mathbf{U}_r)$.

Using this heuristic approach and Remark \ref{rem:Almost-ker-rem}, one can identify anomalies on components of a vector signal, when the absolute value of the scalar product of such vector signal with the left singular vector $\mathbf{u}_L$ is above
certain threshold. The threshold and vector $\mathbf{u}_L$ can be obtained empirically, calibrating the threshold value and the size of $\mathcal{H}_L(\Sigma_N)$ for a signal $\Sigma_\infty$ of the same kind with a known anomaly.

For computational implementation purposes, along the lines of Remark \ref{rem:Almost-ker-rem}, one can define an operation that can be used to identify possible anomalous values in a time series with respect to some training data.

\begin{definition}\label{FADF}
Given $\delta>0$, an integer $L\geq 1$, a times series sample $\Sigma=\{x_t\}_{t=1}^T\subset \Sigma_\infty$, a $(L,\varepsilon)$-PAD $\mathbf{P}$ for $\Sigma_\infty$ based on $\Sigma_N$, and $\mathbf{p}\in \ker(\mathbf{P})$, we write $\phi_\delta(\mathbf{p},x_t)$ to denote the forward anomaly detection function (FADF) determined by the expression
\begin{equation}
\phi_{\delta}(\mathbf{p},x_t) = H_\delta(|\mathbf{p}^\top \mathbf{x}_L(t)|).\label{eq:FADF}
\end{equation}
\end{definition}

One of the advantages of the FADF previously defined, is that it has an easy computational implementation in array/vector oriented programming languages. 

\begin{rem}
Even though, the estimates obtained in this section provide specific criteria to identify possible regions for anomaly localization or behavioral changes, for discrete time signals data that have been measured from some given dynamical system, dealing with real sensor signals corresponding to cyber-physical systems can be very challenging as observed by \cite{DigitaTwins_CPS_9438560}, and may require some heuristic tuning of the parameters and meta-parameters involved in the computational process corresponding to anomaly detection algorithms. This fact has been taken into account for the computational implementation of the ideas presented in this section. As many of the smart-devices for fitness tracking are equipped with sensors and other components, that make them quite responsive to human-application interaction, tuning can also be performed using input data generated by the device interaction with the user.
\end{rem}

\section{Algorithms}\label{sec:algorithms}

We can apply the results in Section \S\ref{sec:anomaliesDetector} to obtain some computational procedures for time series anomaly detection.

\subsection{Time Series Anomaly Detection Algorithm} 
The results and concepts studied in section \S\ref{sec:anomaliesDetector} can be implemented using a prototypical computational procedure described in Algorithm \ref{alg:main_anomaly_alg_3}. For implementation purposes, thinking of array/vector oriented programming languages and computational tools, the prototypical Algorithm \ref{alg:main_anomaly_alg_3} returns vectors $\mathbf{d}(\Sigma)$, $\mathbf{a}(\Sigma)$ in $\mathbb{R}^N$. The vector $\mathbf{d}(\Sigma)$ corresponds to a tagging/identifying array, that can be used to approximately localize anomalous values in $\Sigma$, in terms of the $L$-normality region corresponding to a size $S$ sub-sample $\Sigma_{S}$ collected from a time series sample $\Sigma$, according to Step {\bf 0:} in Algorithm \ref{alg:main_anomaly_alg_3}. Once the identifying vector $\mathbf{d}(\Sigma)$ has been computed, by taking the Hadamard product $\mathbf{d}\odot\mathbf{x}_N(N)$ we can obtain a vector $\mathbf{a}(\Sigma)$ with possible anomalous values located in the vector components corresponding to the components of $\mathbf{d}(\Sigma)$ with values equal to one.

\begin{algorithm2e}[h!]
\caption{{\bf Projective time series anomaly detection algorithm}}
\label{alg:main_anomaly_alg_3}
\SetAlgoLined
 \KwData{$\Sigma=\{x_t\}_{t=1}^T\subset \mathbb{R}$, $1\leq L\leq T-1$, $L+1\leq N\leq T$, $L+1\leq S\leq T$, $\nu>0$, $\delta>0$}
  \KwResult{$\mathbf{d}(\Sigma)$, $\mathbf{a}(\Sigma)$}
\begin{itemize}
\item [0:] Collect sample $\Sigma_S:=\{x_t\}_{t=1}^S$\;
\item [1:] Set $\mathbf{H}=\mathcal{H}_{L}(\Sigma_S)$\;
\item [2:] Compute $\mathbf{p} \in \operatorname{ker}_\nu (\mathbf{H}^\ast)$\;
\item [3:] \For{$j\gets1$ \KwTo $N$}{
    \begin{itemize}
    \item[3.1:] Set $d_j:=\phi_\delta(\mathbf{p},x_j)$\;
    \item[3.2:] Set $\mathbf{d}(\Sigma)=[d_1 \; d_2 \; \cdots \; d_N]^\top$\;
    \item[3.2:] Set $\mathbf{a}(\Sigma)=\mathbf{d}\odot\mathbf{x}_N(N)$\;
    \end{itemize}
    }
\end{itemize}
\KwRet{$\mathbf{d}(\Sigma)$, $\mathbf{a}(\Sigma)$}
\end{algorithm2e}

We can also observe that, in the Step 2 of Algorithm \ref{alg:main_anomaly_alg_3}, we need to compute a vector $\mathbf{p}$ such that $ \mathbf{p} \in \operatorname{ker}_\varepsilon (\mathbf{H})$. For that task, in practice, we consider working with the hermitian matrix $\mathbf{H} \mathbf{H}^{\ast}$. Then, we use two methods to find $ \mathbf{p}$: the first one uses the inverse power method with Rayleigh quotient, described in section \S\ref{sec:preliminaries}, to take advantage of its cubic convergence, and the fact that this method can be used to find one specific vector. The second method is to use optimized numerical routines for hermitian matrices to obtain all the left singular vectors of the matrix $\mathbf{H} \mathbf{H}^{\ast}$ and then choose the singular vector associated to the smallest singular value. Observe that those two approaches are computationally faster than calculating the SVD for the full matrix $\mathbf{H}$, because the size of the matrix is reduced by selecting $ L < S - L + 1$ for the Hankel matrix. The time comparison between all these methods will be shown in the experiments section.

\section{Numerical Experiments}\label{sec:experiments}

We tested Algorithm \ref{alg:main_anomaly_alg_3} with three different signals: one synthetic signal and two real signals that are generated from 
measurements of acceleration done with the Phyphox software (see \cite{staacks2018advanced}). The acceleration is measured using the accelerometer of an Samsung Galaxy A51 cellphone, and the numerical experiments were done using: Python 3.9.12, NumPy 1.22.3 and SciPy 1.8.0 on a PC running Linux Mint 20.2 Uma 
with an Intel Core i7 8700 with 6 cores and a G.Skill RipJaws Serie V 8 GB DDR4 2400. For these experiments, the parameters involved were tuned 
manually, using also the \texttt{anomaly\_detector\_calibration\_widget.ipynb} widget, until the anomaly is properly identified without identifying non anomalous regions.

\subsection{Example 1}
\begin{figure}[h!]
	\centering
	\includegraphics[width=0.48\textwidth]{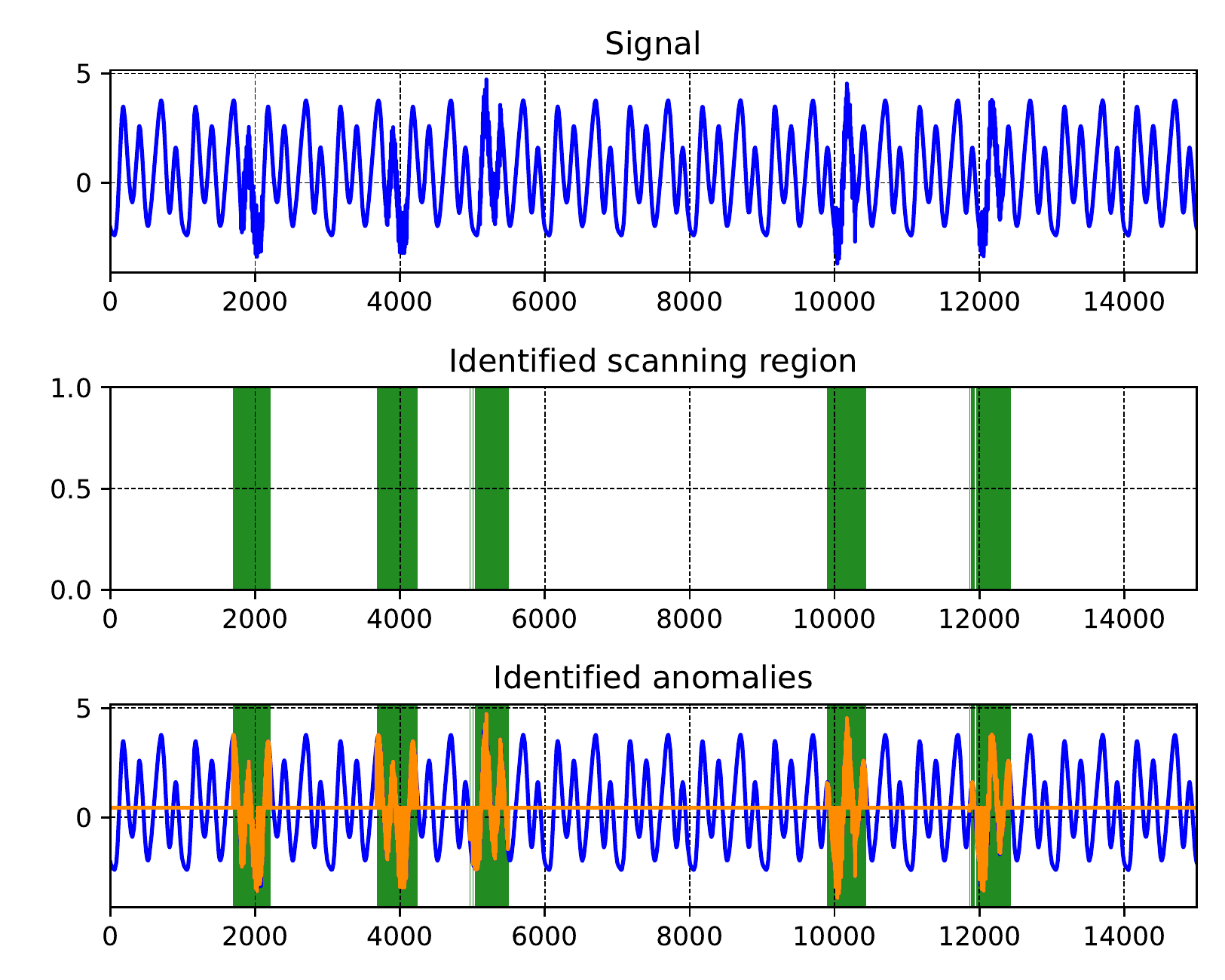}
	\caption{Synthetic signal with $L = 150$, $S$ = 1200, tolerance = 1.2 and using the SVD method}
	\label{fig:synthetic_signal_anomalies_SVD}
\end{figure}

The synthetic signal of this example is a linear combination of sines and cosines with varying frequency, with some localized pseudo-random noise on certain regions to generate anomalies. This signal is shown in Fig. \ref{fig:synthetic_signal_anomalies_SVD} and the data can be found on the file \texttt{synthetic\_signal.csv.}

It can be seen in Fig. \ref{fig:synthetic_signal_anomalies_SVD} that the PAD correctly identifies the region where the signal has been disturbed with the random noise.

\subsection{Example 2}
\begin{figure}[h!]
	\centering
	\includegraphics[width=0.48\textwidth]{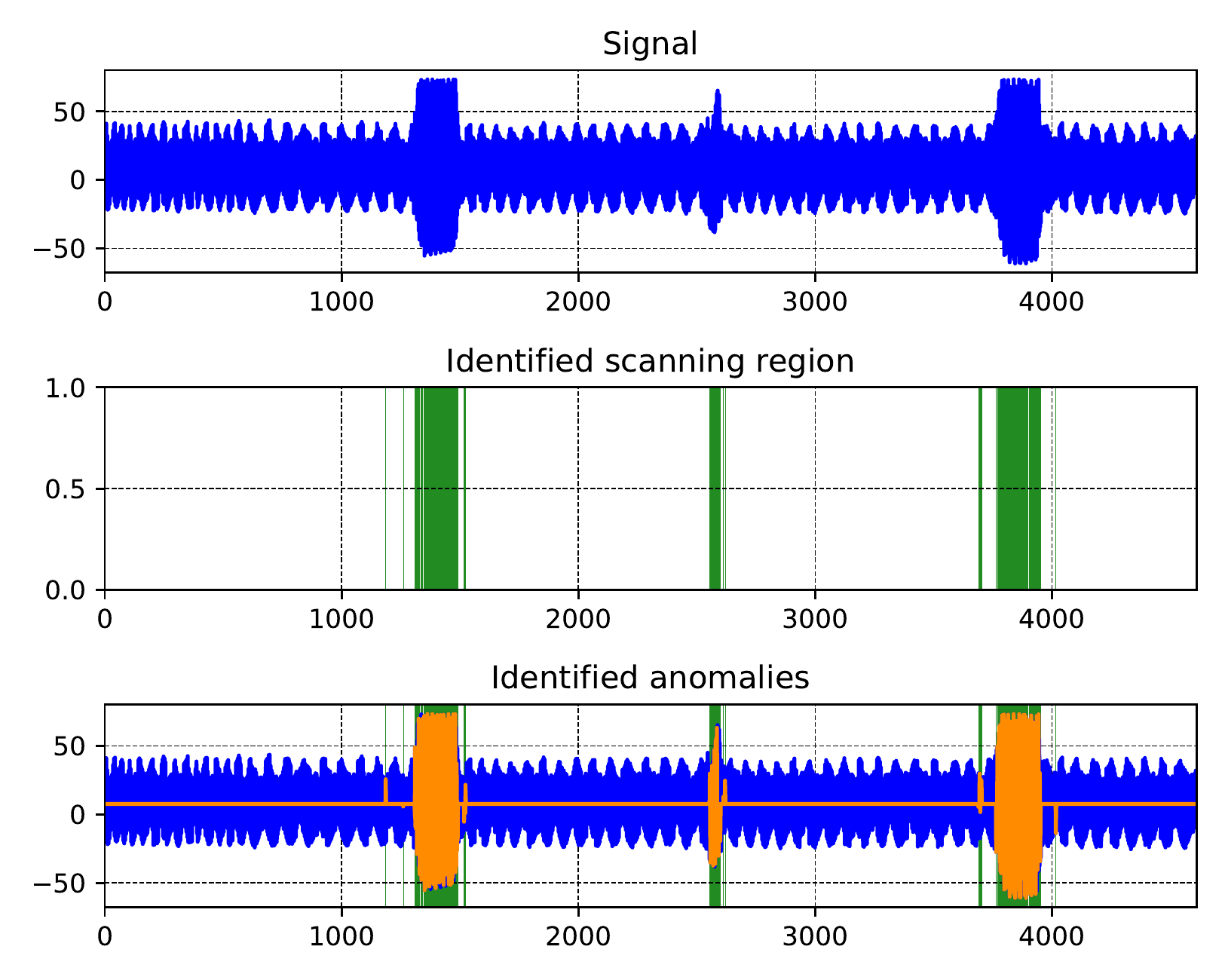}
	\caption{Signal from electric back massager with its identified anomalies with $L = 75$, $S$ = 1300, tolerance = 0.4 and using SVD method}
	\label{fig:real_signal_1_anomalies_SVD}
\end{figure}

This second example corresponds to a real signal obtained from an electric back massager with continuous variable intensity. The massager was set on the lowest intensity, and to generate the anomaly, the intensity was suddenly 
increased and then set back to lowest intensity. The signal is shown in Fig. \ref{fig:real_signal_1_anomalies_SVD} and the data for the signal is available in the file \texttt{real\_signal\_1.csv}. We can see in Fig. \ref{fig:real_signal_1_anomalies_SVD} the anomalies and the scanning region of the anomalies. It can be observed that the anomalies are successfully identified with the manually calibrated parameters $L$, $N$ and tolerance. The results for this signal can be replicated with the script \texttt{example\_real\_signal\_1.py}

\subsection{Example 3}
\begin{figure}[h!]
	\centering
	\includegraphics[width=0.48\textwidth]{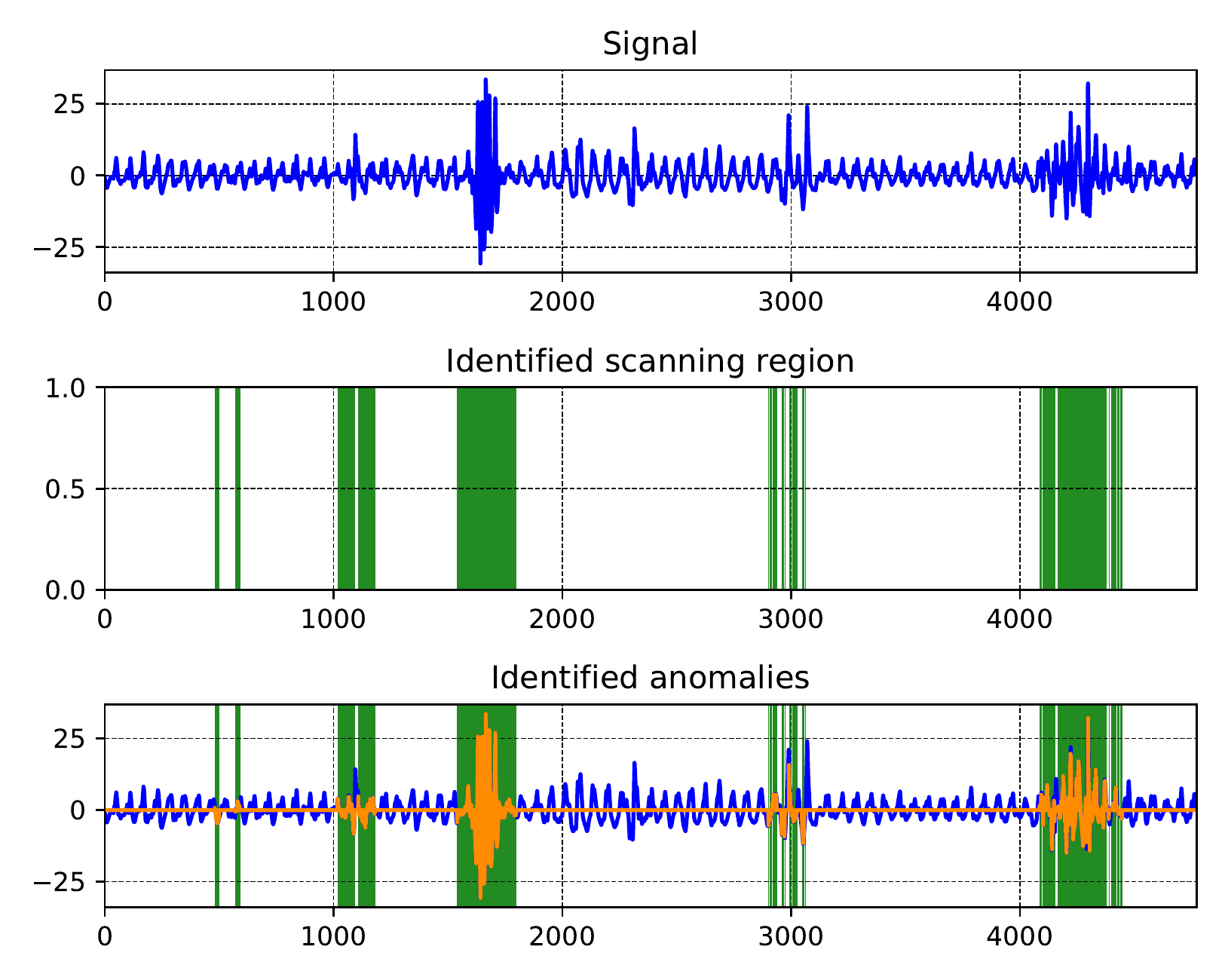}
	\caption{Signal from walking with its identified anomalies  with $L = 60$, $S$ = 900 and tolerance = 1.01 and using Inverse Power Method}
	\label{fig:real_signal_2_anomalies_IPM}
\end{figure}

This third example corresponds to a real signal generated by one of the authors walking, keeping a steady rate and suddenly jumping and changing pace to produce anomalies. The signal is shown in Fig. \ref{fig:real_signal_2_anomalies_IPM} and the data for the signal is available in the file \texttt{real\_signal\_2.csv}. It can be seen in Fig. \ref{fig:real_signal_2_anomalies_IPM} that walking interruptions and abrupt changes in pace, are succesfully identified by the PAD.
The results for this signal can be replicated with the script \texttt{example\_real\_signal\_2.py}

\subsection{Example 4: Using different tolerances}
\begin{figure}[h!]
	\centering
	\includegraphics[width=0.48\textwidth]{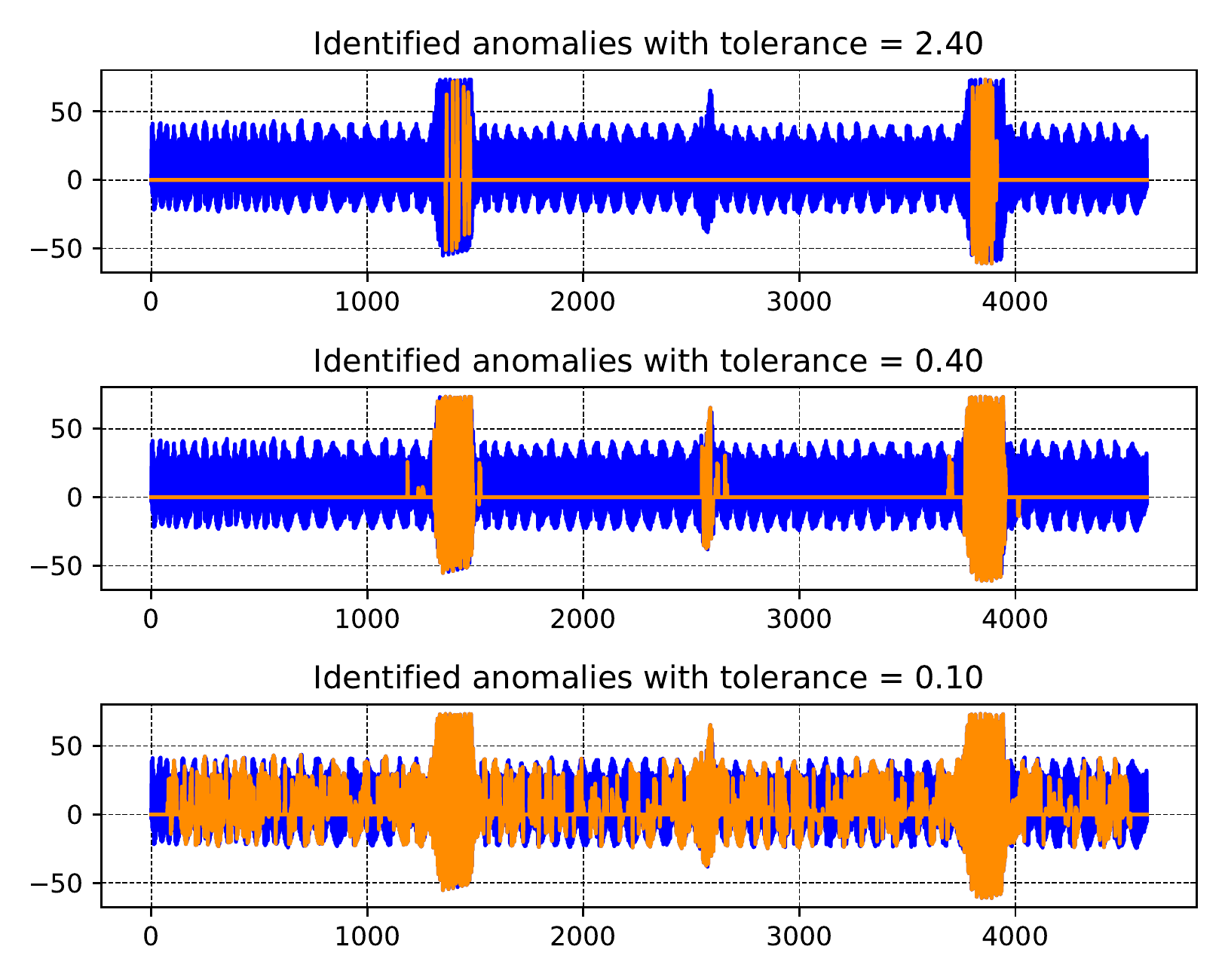}
	\caption{Identified anomalies for different tolerance for the first real signal, with $L= 60$ and $S = 900$}
	\label{fig:real_signal_1_different_tolerances}
\end{figure}

We can see the effect of having different thresholds for the first real signal in Fig. \ref{fig:real_signal_1_different_tolerances}. Observe that with different values for the tolerance, the anomalies 
may not be properly identified, or falsely identify 
non anomalous regions. The higher the tolerance is, the more anomalies might not be identified and pass as 
false negatives. And if the tolerance too low, some segments of the signal are identified as false anomalies. The results corresponding to this experiment can be replicated with the script \texttt{example\_different\_tolerance.py} 

\subsection{Time to get the vector $\mathbf{p}$ with different methods}

To quantify what method computes a $\mathbf{p}$ faster, on Step 2 of Algorithm \ref{alg:main_anomaly_alg_3}, we ran 600 trials for each signal with each method and took the average time for each signal and method. The $L$ and $S$ used for each signal are the ones shown in Fig. \ref{fig:synthetic_signal_anomalies_SVD}, \ref{fig:real_signal_1_anomalies_SVD} and \ref{fig:real_signal_2_anomalies_IPM}, respectively. Then, in Table \ref{table:times_for_each_method}, we can see that the fastest method in average for all signals is the inverse power method if the lowest singular value is approximated beforehand. The next fastest method is obtained using the hermitian eigenvector solver \texttt{numpy.linalg.eigh} 
with the matrix $\mathbf{H} \mathbf{H}^{\ast}$. Then, the second slowest we get is the \texttt{numpy.linalg.svd} function with the matrix $\mathbf{H} \mathbf{H}^{\ast}$, and the slowest method is obtained using the \texttt{numpy.linalg.svd} function with $\mathbf{H}$. The time calculating $\mathbf{H} \mathbf{H}^{\ast}$ is considered in each method, except for the SVD method that uses $\mathbf{H}$.
The times are obtained with the \texttt{time\_measurements.ipynb} notebook.

\begin{table}
\begin{center}
\caption{Average time for each method in seconds}
\label{table:times_for_each_method}
\begin{tabular}{lrrrr} Signals & IPM & \texttt{eigh} & \texttt{svd}($\mathbf{H H^{\ast}}) $ & \texttt{svd}($\mathbf{H}$) \\ \hline
Massager & 0.001256 & 0.001793 & 0.002407 & 0.008883 \\ Walking & 0.000946 & 0.001272 & 0.001608 & 0.005264 \\ Synthetic & 0.027404 & 0.028571 & 0.052334 & 0.097148 \\ \hline \end{tabular}
\end{center}
\end{table}

\section{Hints and Future Work}
\label{sec:Future}
Based on the theoretical and computational elements considered previously, the subspace method presented in this document provides an effective tool for anomaly detection at a relatively low computational cost, specially when implemented using iterative techniques to identify elements in the approximate kernel of the Hankel-type matrix corresponding to the training data. In order to improve efficiency and precision, we will work on lower level implementations of some of the key routines, like the computation of elements in the previously considered approximate kernels. We will also work on more adaptive and artificially intelligent schemes that combine linear and nonlinear models, along the lines of \cite{FVides_SPAAR}, where the linear part of the models can be computed using the methods discussed in this document. Applications to digital twins simulation, aimed to provide alternative solution strategies to the problems identified in \cite{DigitaTwins_CPS_9438560}, will also be studied and discussed in future communications.

\section{Conclusion}
The subspace method for anomaly detection proposed in this section, together with some of its variants that have been implemented using the programs in \cite{GIT_Project}, have proved to be effective at a relatively low computational cost, and these features make our method suitable for implementation on fitness and health tracking smart devices, with limited computational resources, and because of the relative simplicity of the numerical implementations of our subspace method, our techniques are also suitable for artificially intelligent collaborative schemes, that can be used to improve anomaly detection precision, when more computational resources become available, for instance, by using smart-bands connected to smartphones and eventually to the cloud.

\section{Data Availability}

The programs and data that support the findings of this study are openly available in the PAD repository, reference number \cite{GIT_Project}.

\begin{ack}
Some of the structure preserving computations corresponding to the numerical experiments documented in this paper were performed using the computational resources of the Scientific Computing Innovation Center of UNAH, as part of the researh projects PI-063-DICIHT and PI-174-DICIHT. The work of Esteban Segura was partially supported by Escuela de Matem\'atica and
CIMPA, Universidad de Costa Rica, Costa Rica, through the project 821-B7-254.
\end{ack}

\bibliography{NeuralNOR}       
\end{document}